\documentstyle[12pt]{article}
\begin{document}
\newcommand{\ket}[1] {\mbox{$ \vert #1 \rangle $}}
\newcommand{\bra}[1] {\mbox{$ \langle #1 \vert $}}
\newcommand{\bkn}[1] {\mbox{$ < #1 > $}}
\newcommand{\bk}[1] {\mbox{$ \langle #1 \rangle $}}
\newcommand{\scal}[2]{\mbox{$ \langle #1 \vert #2  \rangle $}}
\newcommand{\expect}[3] {\mbox{$ \bra{#1} #2 \ket{#3} $}}
\newcommand{\ki}{\mbox{$ \ket{\psi_i} $}}
\newcommand{\bi}{\mbox{$ \bra{\psi_i} $}}
\newcommand{\p} \prime
\newcommand{\e} \epsilon
\newcommand{\la} \lambda
\newcommand{\om} \omega   \newcommand{\Om} \Omega
\newcommand{\cc}{\mbox{$\cal C $}}
\newcommand{\w} {\hbox{ weak }}
\newcommand{\al} \alpha
\newcommand{\bt} \beta
\newcommand{\be} {\begin{equation}}
\newcommand{\ee} {\end{equation}}
\newcommand{\ba} {\begin{eqnarray}}
\newcommand{\ea} {\end{eqnarray}}

\def\lrD{\mathrel{{\cal D}\kern-1.em\raise1.75ex\hbox{$\leftrightarrow$}}}

\def\lr #1{\mathrel{#1\kern-1.25em\raise1.75ex\hbox{$\leftrightarrow$}}}

\overfullrule=0pt \def\sqr#1#2{{\vcenter{\vbox{\hrule height.#2pt
          \hbox{\vrule width.#2pt height#1pt \kern#1pt
           \vrule width.#2pt}
           \hrule height.#2pt}}}}
\def\square{\mathchoice\sqr68\sqr68\sqr{4.2}6\sqr{3}6} 
\def\lrpartial{\mathrel
{\partial\kern-.75em\raise1.75ex\hbox{$\leftrightarrow$}}}

\begin{flushright}
October 30th, 1997
\\\end{flushright}
\vskip 1. truecm
\vskip 1. truecm
\centerline{\LARGE\bf{The notions of time and evolution}}
\vskip 2 truemm
\centerline{\LARGE\bf{in quantum cosmology}}
\vskip 1. truecm
\vskip 1. truecm

\centerline{{\bf R. Parentani}}
\vskip 5 truemm

\centerline{
Laboratoire de Math\'ematiques et Physique Th\'eorique,
CNRS UPRES A 6083}
\centerline{Facult\' e des Sciences, Universit\'e de Tours, 37200 Tours, France}
\centerline{
e-mail: parenta@celfi.phys.univ-tours.fr}
\vskip 5 truemm
\vskip 1.5 truecm
\vskip 1.5 truecm

{\bf Abstract }\\
We re-examine the notions of time and evolution
in  the light of the mathematical
properties of the solutions of the Wheeler-DeWitt equation
which are revealed by an extended adiabatic treatment.
The main advantage of this treatment is to organize 
the solutions in series that make explicit both the 
connections with the corresponding Schr\"odinger equation as well as
the modifications introduced by the quantum character of gravity. 
When the universe is macroscopic, the ordered character
of the expansion leads to connections with the 
Schr\"odinger equation so precise that the interpretation
of the solutions of the Wheeler-DeWitt equation is unequivocally
determined. On the contrary, when the expansion
behaves quantum mechanically, i.e. in the presence of backscattering,
 major difficulties concerning the interpretation persist.

\vfill \newpage

\section{Introduction}

It is now well known that in quantum cosmology, when working
with the solutions of the Wheeler-DeWitt (WDW) equation, one
completely looses the notion of time, i.e. the notion of an {\it external} parameter that
clocks events. Indeed, the WDW constraint equation 
implies that the universe is in an eigenstate of zero total energy\cite{hartle}-\cite{kiefer}. 
In a Schr\"odinger sense, this means that one works with a stationary
state. 

Before investigating the problem of the description of evolution in
quantum cosmology, it should be noted that in classical cosmology,
when working with the action in the Hamilton-Jacobi formalism, one also looses
the notion of the (Newtonian) time. As in quantum cosmology,
this results from the invariance of the theory under arbitrary reparametrizations
of time. It should also be noticed that a similar 
disappearance applies to space as well, in virtue of 
the full reparametrization invariance of general relativity. 
Surprisingly, this problem
has attracted much less attention
that the one associated with time, 
see however \cite{HH}\cite{wdwgf}\cite{wdwpt}.
In any case, when 
analyzing the solutions of the Wheeler-DeWitt equation, 
 one must keep in mind
that certain properties 
might posses an equivalent
version in classical cosmology and, more importantly,
one should sort out
these classical aspects from those which are specific to quantum gravity.
These concerns are rarely found in the literature, 
see however \cite{salopek}\cite{barbour}\cite{bprmp}.

The absence of time in quantum cosmology leads at least to three questions:

\noindent
1. Is the notion of time intrinsic to the notion of (quantum) evolution ?

\noindent
2. How evolution should be described in the absence of time ?

\noindent
3. When and why this new description coincides with the usual one
 based on a Schr\"odinger equation ?

That time might not be necessary in cosmology is not really a surprise.
For instance, cosmological events are delivered to astronomers 
ordered in terms of their red shifts $
z= a_{reception}/ a_{emission} -1 $ which bear no direct information about 
lapses of proper time $\Delta t$
 between emission and reception. Indeed, the determination of 
the latters requires the knowledge of the
expansion rate $\dot a(a) = da/dt$ from $a_{emission}$ until $a_{reception}$.
Thus, at first sight, since one has lost the notion
of time, it seems that 
this rate is not physically meaningful.
However, 
it {\it becomes} 
meaningful when it 
is {\it compared} to rates governing 
local processes, such as growth of perturbations.
This is because these rates 
issue from evolution laws governed by lapses of (proper) time. 

This discussion brings about an important concept that we shall use throughout in
this paper: it is the comparison of the rates characterizing two 
different processes that leads, 
a posteriori, to the notion of time. This is why we shall 
analyze the transitions amplitudes of heavy atoms
induced by absorption or emission of photons that occur in an expanding universe.
Notice 
that the physical necessity of comparing two rates to re-introduce the notion 
of time is reminiscent of the description of 
planar orbits of an 
isolated system in Newtonian mechanics.
In that case, one can describe intrinsically, i.e. without reference to an
{\it external} time, the trajectory in the form $r(\theta)$. 
However, to give physical sense
to the ``extrinsic'' description $r(t), \theta(t)$, one needs an additional 
system to tell what $t$ means\cite{barbour}\cite{bprmp}\cite{mtw}.

To further discuss the issues raised by the three questions, one must consider
 technical aspects more closely. 
Firstly, we shall pursue the discussion in minisuperspace. In this 
restricted space, classical cosmology is described by the history of 
the scale factor $a(t)$. The main advantage of this truncation is that it leads to explicit
equations whose solutions can be fully analyzed. 
The physical relevance of this radical truncation is greatly enlarged
once it is conceived as the first order term in an expansion in local
gravitational deformations\cite{HH}\cite{salopek}.  In particular,
in such an expansion,
it is consistent to consider quantum matter fields carrying non vanishing 
momenta evolving in homogeneous three geometries, thereby
allowing the study of local phenomena\cite{wdwpt}. 
When mini-superspace is conceived in this manner,
the rather symmetrical roles played by $a$ and $\phi$ 
(the homogeneous part of a matter field) 
in a strict mini-superspace reduction
become clearly distinguishable 
since the local perturbations of all fields (matter and gravity) are 
minimally coupled to $a$ 
 and not to $\phi$. 

Secondly, of great importance, is the
choice of the formalism used to investigate the solutions of the WDW equation.
Indeed, the choice of the formalism plays a double role.
First, it organizes the mathematical approximations  
necessary to construct non-trivial solutions. Secondly,
it partially determines the physical interpretation of these solutions.
In fact, a careful examination of the
literature reveals that authors who use very different mathematical expansions
reach generally different conclusions concerning the 
interpretative aspects, see \cite{hartle}-\cite{kiefer},\cite{wdwin}.
In any case, at present, no 
fully consistent interpretation of the 
solutions of the WDW equation has been reached.
At the end of this paper, we shall present the reasons that prevent a simple 
interpretation of these solutions.
However, in spite of these difficulties, unambiguous answers 
to the three questions raised above can be obtained in certain cases. 
To this end, we shall apply, following \cite{wdwpc}, 
an (extended) adiabatic treatment to the WDW equation.
The merits of this treatment are the following.

First, when applied both to the Schr\"odinger and the WDW equations describing 
the same quantum processes,
this treatment makes the comparison of both descriptions of the transitions
quite transparent. This is because, in both descriptions,
the dynamical role of the expansion rate of the universe is made explicit.
Indeed, non-adiabatic transitions are directly induced by the expansion.
This is to be opposed to time dependent perturbation theory wherein
transitions are induced by the interaction hamiltonian and
bear therefore no direct relation to cosmology. 
Notice also that
 in the absence of adiabatic transitions, there is nothing to clock.
Thus, time cannot be recovered from a universe 
in which matter is in an eigenstate
(or a frozen superposition thereof).
In fact, we shall recover the notion of time
from the transition rates induced by the expansion.
In this, the analysis of non-adiabatic transitions precisely formalizes
the idea expressed above following which time reappears from a comparison of the rates 
of two physical processes. Indeed, no reference to external properties, like 
the trajectory of the ``peak''\cite{Hal} of the wave function of the universe, will be used.

Moreover, the adiabatic treatment leads to an {\it exact} rewriting of the WDW equation
in which the consequences of the dynamical character of gravity are displayed.
It does not, in itself, introduce any kind of approximations nor does it
require specific matter properties. Instead, it
organizes the solutions of the WDW equation into series characterized by adiabatic 
parameters. 
Then, the validity of a {\it truncation} of these series requires that the parameters, say,
 be small. This in turn puts physical restrictions on the space of
valid truncated solutions.
The important fact is that this organization of the series 
answers {\it by itself} question 3, i.e. it both determines the 
precise conditions that guarantee that the classical and quantum descriptions coincide
and explicitizes the mechanisms by which the two descriptions differ
when the conditions are not met. 
In brief, the main condition that validates a
truncation of the series is that the universe be macroscopic, i.e. that the matter
sources driving gravity be macroscopic\cite{HH}\cite{vil}\cite{BV}. (Recall
that the appropriate character of the adiabatic treatment follows from
the fact that one deals with dynamically coupled systems
which are characterized by very different time-energy scales.)
Then, the truncated adiabatic treatment
shows that microscopic quantum transitions
evolve according to a unitary evolution in the mean geometry parametrized by $a$, 
in a manner similar that non-adiabatic electronic (light) transitions occurring in molecules
can be parametrized in terms of the (heavy) nuclei positions.
Indeed, in cosmology, the rest mass of all matter delivers through the WDW constraint a 
kind of (macroscopic) inertia to gravity which fixes the geometry in which the 
(microscopic) transitions take place.
Notice that the value of the Planck mass plays no role into this division in macroscopic
and microscopic energy scales. We emphasize this point: for non empty macroscopic
universes, it is unappropriate to develop the solutions of the WDW equation in series 
of the Planck mass, contrary to what is adopted in the ``standard treatment'' presented 
for instance in \cite{kiefer}.

Finally,  the adiabatic treatment allows to question the problem
of the interpretation of the solutions of the WDW equation in well 
defined mathematical terms. In 
the case of macroscopic universes,
the coefficients that must be interpreted as the amplitudes to find
the n-th state at $a$ are designated by the formalism. As pointed out to me
by S. Massar, this procedure to reach the interpretation of the WDW solutions 
bears many similarities to the original reasoning that Max Born used to 
reach the probabilistic interpretation of the solutions of the Schr\"odinger 
equation\cite{born}.
In both cases, it is through an examination of the mathematical properties of the
solutions describing quantum transitions 
that the physical interpretation is, a posteriori, reached.

The properties and the 
inferred interpretation are the following\cite{wdwin, wdwpc}. 
As long as matter is close to equilibrium and 
evolving in a macroscopic universe, the evolution extracted from the WDW equation
coincides with the Schr\"odingerian evolution of the corresponding problem.
The identification 
of the amplitude to find matter in the 
n-th state is unambiguous and the physical interpretation follows from this identification.
When matter is far from equilibrium but the universe still
macroscopic, the evolution differs 
from the Schr\"odingerian one, 
even though it is still unitary.
As before, the identification and the interpretation of the amplitudes 
evolving unitarily are unambiguous. 
In both cases, unitarity follows from the ordered semi-classical expansion of 
the macroscopic universe.

Therefore, there is a major difficulty when the cosmological 
expansion can no longer be described semi-classically.
In this case indeed, through backscattering, 
one obtains non-vanishing coupling terms between
quantum matter states associated with expanding and contracting universes.
The present difficulties in interpreting the solutions of the WDW
equation stem from the consequences of these coupling terms.
Most probably, they require an extension of the 
usual concepts prevaling in quantum mechanics.
We shall conclude this paper by presenting and commenting various approaches 
that have been proposed to overcome these difficulties.


\section{The adiabatic treatment applied to quantum mechanics in 
classical cosmology}\label{adia}

When studying quantum processes in classical cosmology, the
expansion of the universe is treated at the background field approximation (BFA),
i.e. $a= a(t)$ is given from the outset and thus unmodified by the quantum transitions
that one investigates. This is of course an approximation since, in general relativity,
 gravity is coupled to all forms of energy. To take into account this coupling is one of 
the jobs of quantum gravity.

In classical cosmology, the point that is crucial for us is 
that the expansion law, $a= a(t)$, leads, 
in the general case, to time dependent hamiltonians. Indeed only degenerate cases,
like purely photonic homogeneous universes, are characterized by constants of motion.
Therefore, in general, one deals with the (explicitly) time dependent 
Schr\"odinger equation
\be
i \partial_t \ket{\psi(t)} = H(t) \ket{\psi(t)}
\label{sch}\ee
(In this equation, as everywhere in this paper, we designate by $t$ the 
proper time evaluated in the classical universe one is dealing with.)
The main consequence of eq. (\ref{sch}) is that there will be no stationary states,
a very useful feature on which we shall base our investigation of the role of time 
in cosmology.
Physically it means that the expansion rate $\dot a $
induces quantum transitions. 
In this respect it should be noticed that the whole
time dependence of $H(t)$ comes from $ a(t)$ only, i.e.
\be
H(t) = H(a(t))
\label{ha}
\ee
There is no {\it external} time dependence.
In other words, we postulate that the universe is isolated.
This possesses some flavor of General Relativity and will be automatically 
implemented when working in quantum cosmology.

The adiabatic treatment consists in developing the solutions of eq. (\ref{sch})
in terms of instantaneous (normalized) eigenstates of $H(a)$
\be
H(a)\ket{\psi_n(a)} = E_n(a) \ket{\psi_n(a)}
\label{insteig}
\ee
\be
\scal{\psi_n(a)}{\psi_m(a)}= \delta_{n,m}
\label{normal}\ee
One can already see the appropriateness of this treatment.
It naturally incorporates the fact that the eigenstates and their eigenvalues
depends on $t$ through $a(t)$ only. This is exactly like the red-shift mentioned in the
introduction: the energy of photons measured in proper time scales as 
\be
\om(a) = \om_{emission} { a_{emission} \over a }
\label{emiss}
\ee
which is a special case of eq. (\ref{insteig}).

By factorizing, as in time 
dependent perturbation theory\cite{LL}, the ``free'' kinetic factor $exp(-i\int^t dt' E_n(t'))$,
 the development in this basis reads
\be
\ket{\psi(t)} = \sum_n c_n(t) \ e^{-i\int^t dt' E_n(t') }
\ket{\psi_n(a(t))}
\label{dev}
\ee
By inserting this development into eq. (\ref{sch}),
one obtains the equation that determines
 the time dependence of the coefficients $c_n(t)$:
\be
\partial_t c_n = \sum_{m\neq n} 
\scal{\partial_t\psi_m(t)}{ \psi_n(t)}\ exp\left[{-i\int^t dt'
( E_m(t') - E_n(t') )}\right] c_m(t)
\label{eqtcn}
\ee 
It is instructive to compare this equation with the one obtained in time dependent 
perturbation theory. In that case, the matrix elements 
$\scal{m}{V \vert n}$ of the
perturbation $V$ induce transitions among the ``free'' states $\ket{n}$ and $\ket{m}$.
Here, these matrix elements are replaced by $\scal{\partial_t\psi_m(t)}{ \psi_n(t)}$.
Simple algebra gives
\ba
\scal{\psi_m}{\partial_t\psi_n}
&=&
{\bra{\psi_m} \partial_t H \ket{\psi_n}\over E_n - E_m }\quad , \quad n \neq m
\label{dHdt}
\ea
Thus, it is the time dependence of the instantaneous eigenstates, which is 
induced and determined by the time dependence of $H$, 
that induces, in turn, transitions among these states.

Therefore, in classical cosmology, non-adiabatic transitions 
are caused by the expansion law $a(t)$ since
$H(t)$ depends on time through $a$ only.
Moreover, one can rewrite eq. (\ref{eqtcn}) directly in terms of $a$
as
\be
\partial_a c_n = \sum
_{m\neq n} 
\scal{\partial_a\psi_m(a)}{ \psi_n(a)}\  
exp\left[{-i\int^a da' {1 \over \dot a(a')}(E_m(a') - E_n(a'))}\right]  c_m(a)
\label{eqtcn2}
\ee
One sees that the only place where time appears is in the difference of phases
between neighboring states. Furthermore it appears parametrized by $a$,
through the inverse rate $1/\dot a(a)$. 
Given this rate, 
the lapse of proper time from $a_0$ is of course given by
\be
\Delta t(a) = \int^a_{a_0} \!da' \;{ 1 \over \dot a(a')}
\label{dota}
\ee
This equation makes explicit the fact 
that $\dot a(a)$ should be known to convert
Doppler shifts controlled 
by $a_{reception}/ a_{emission}$ into proper time lapses,
{\it c.f.} the Introduction.

As such eq. (\ref{eqtcn2}) is simply a rewriting of the Schr\"odinger equation in which the role 
of the expansion rate has been put forward. 
However, its real usefulness will become manifest in 
Section 3 since it is precisely in that guise
that the evolution of the $c_n$, i.e. the weights of the instantaneous eigenstates,
are delivered in quantum cosmology.
This is not an accident, as we explain below.
Moreover,
eqs. (\ref{eqtcn}, \ref{eqtcn2})  prepare the analysis of the solutions
both from the mathematical and the physical point of view. 

We start with the physical point of view. 
As noticed in the Introduction, Max Born was lead to the interpretation
of the wave function by the properties of first order transition amplitudes.
More precisely, the three following points 
are such that, according to him, ``only one interpretation is possible''\cite{born}:
1. When initially $c_m = \delta(m,n)$, 
this means that the system is
in the n-th state characterized by the eigen-energy $E_n$,
2. Mathematically, the final
values of $c_m$ are asymptotically constant and satisfy $\sum_m \vert c_m \vert^2 = 1$ (for 
all hermitian hamiltonians and for normalized eigenstates)
and 3. Experimentally, the asymptotic ``electron'' is found 
in {\it one} of the outcomes labeled by $n$.
These three properties are also found in the adiabatic treatment
of $\ket{\psi(t)}$. This is not surprising in view of the analogies with perturbation theory.
Thus, the properties of $c_n(t)$, solutions of eq. (\ref{eqtcn}), can also be used to reach the
interpretation of $\ket{\psi(t)}$ along Born's lines. 
Moreover, similar properties will be found as well in quantum cosmology
when an extended adiabatic treatment is applied 
to the WDW equation. We shall then base our
interpretation of the wave function of the universe on the properties
so obtained.

From a mathematical point of view, the first useful property of the adiabatic treatment
follows from the fact that the instantaneous eigenstates form
 a ``comoving'' vector basis in Fock space.
By comoving, we mean that no transition among eigenstates occurs
 in the limit in which 
the characteristic time of the expansion (i.e. $ a/\dot a$) is much larger than the 
characteristic time of quantum transitions (i.e. the time for the Golden Rule to be valid).
To work in the adiabatic approximation simply means that one neglects
completely these transitions.
Moreover, to first order in the non-adiabaticity, the probability to find 
a transition takes a universal form controlled by an exponentially small factor, like
in a tunneling process, see \cite{wdwpc}. 

The second useful property concerns the possibility of enlarging the dynamics.
Up to now indeed, the adiabaticity concerned only the quantum dynamics of the (light)
matter degrees of freedom since $a(t)$ was treated at the BFA.
However, the adiabatic treatment is naturally enlarged so as to 
take into account the quantum dynamics of the (heavy) degree $a$ as well.
This ``extendible'' property of the adiabatic treatment is precisely what we 
need to investigate (light) transitions in quantum cosmology. 
Moreover, this treatment leads to a rewriting of  
the WDW equation such that the comparison to the 
Schr\"odinger equation is greatly facilitized.
Indeed, the description of light transitions in quantum cosmology 
coincides with the BFA description
when a first order (light) change is applied to the heavy WKB dynamics.
To establish how these mathematical properties are precisely implemented in
the formalism 
is the first aim of next Section.

\section{The extended adiabatic treatment applied to quantum cosmology}\label{adiab}

As we explained, we shall use an  extended adiabatic treatment
as a guide to first identify and then to interpret the coefficient
${\cal{C}}_n(a)$, i.e. the weight of the n-th adiabatic state,
that replaces the coefficient $c_n(a)$ of eq. (\ref{eqtcn2}). 
Before accomplishing this program, 
we briefly present the kinematical properties at work in quantum cosmology
when the eigenstates are stationary, i.e.
in the absence of transitions.
It should be emphasized that this kinematical analysis is preparatory in character
since it reveals the framework in which transitions take place upon considering 
non-degenerate cases. In those
cases only, one can obtain a meaningful notion of evolution based 
on physical processes.

\vskip.5 truecm
{\bf{The (preparatory) notion of propagation in absence of transitions}}
\vskip.2 truecm

\noindent
To work with 
matter systems such that no transition occurs,
requires that the matter states 
be {\it stationary} eigenstates of the hamiltonian $H(a)$ 
\ba
H(a) \ket{\psi_n} &=& E_n(a) \ket{\psi_n} 
\nonumber\\
\partial_a \ket{\psi_n}&=&  0
\label{one}
\ea
In a Schr\"odinger context, this degenerate case would lead to no evolution
in the sense that the coefficients $c_n$ would be constant, see eq. (\ref{eqtcn2}). 
One simply obtains a (frozen) superposition of eigenstates whose relative phases 
depend on time through their eigenvalues. To give a physical substance to these
phases requires either the addition of internal interactions or a coupling to the 
external world, see \cite{wdwin}.

In general relativity restricted to 
minisuperspace, when matter is characterized by an
energy $E_n(a)$, the gravitational action 
satisfies the Hamilton-Jacobi constraint
\ba
H_G(a) + E_n(a)=
{ -G^2  ( \partial_a S_G (a) )^2
 + \kappa a^2 + \Lambda a^4 \over 2 Ga } +
 E_n(a) =0
\label{M12}
\ea
where $G$ is Newton's constant, $\kappa$
is equal to $\pm 1$ or $0$ for respectively open,
closed and flat three surfaces and $\Lambda $ is the
cosmological constant. The
solution of this equation
is simply $S_n (a)
= \int^a da' p_n(a')$ where the momentum of $a$
driven by $E_n(a)$ is
\be
p_n(a) = \mp G^{-1} \sqrt{ 
\kappa a^2 + \Lambda a^4 + 2 Ga E_n(a) }
\label{M8}
\ee
The sign $- \ (+) $ corresponds respectively to an expanding 
(contracting) universe.

Upon working in quantum cosmology, the Hamilton-Jacobi constraint
becomes 
the WDW equation
\be
\left[
H_G(a) + H(a) \right] \ket{\Xi ( a )} =0
\label{wdw}
\ee
The 
matter states 
are still the stationary 
eigenstates of $H(a)$ given in eq. (\ref{one}).
Therefore, as in Schr\"odinger case, the wave function 
$\Xi ( a, \phi )$ can be decomposed as
\be
 \Xi ( a, \phi )= \scal{\phi}{\Xi (a)} = \sum_{n} {\cal{C}}_{n}\;
\Psi ( a; n )\; \scal{\phi}{{\psi_n}}
\label{M13}
\ee
where the weights ${\cal{C}}_{n}$ are constant and 
where the gravitational waves entangled
to their corresponding matter state are solutions of 
\be
\left[
G^2 
\partial^2 _a  
+ \kappa a^2 + \Lambda a^4 + 2 Ga E_n (a) \right]
 \Psi ( a; n ) =0
\label{M14}
\ee
Being second order in $\partial _a $, each equation has two 
independent solutions. 
This has to be the case since classically we can work
either with expanding or contracting universes.
Indeed one verifies that the WKB waves 
\be
 \Psi_{W\!K\!B} ( a; n ) = { e^{ i\!\int^{a}
p_n(a')  da' } \over \sqrt{2 \vert p_n(a) \vert }}
\label{M144}
\ee
 with positive (negative) Wronskians
\be
 W_n = \Psi^*( a; n)\ 
i\!\!\lr{\partial_{a}} \Psi ( a;n ) 
\label{M1444}
\ee
correspond to expanding (contracting)
universes in this semiclassical limit.
It is now through the sign of the
Wronskian rather than at the classical level 
that one can still choose to work
either with expanding or with contracting universes (at least far from a turning point).
Thus, upon abandoning the WKB approximation, one must deal with
superpositions of contracting and expanding universes. As we shall
see later in this article, this mixing
leads to major difficulties concerning the notion of evolution.
 
Before examining transitions, two aspects should be analyzed.
First, we shall construct the Feynman kernel to go from $a_0, \phi_0$ to $a, \phi$
in order to make contact with the notion of free evolution and with the classical
theory.
Secondly, we shall express the exact solutions of eq. (\ref{M14}) in terms of the WKB
solutions eq. (\ref{M144}).
Both aspects shall be exploited upon studying non-adiabatic transitions
in quantum cosmology.

The kernel to go from $a_0, \phi_0$ to $a, \phi$ can be decomposed, as usual, with the
help of the quantum conserved number $n$ according to
\be
{\cal{K}}(a, \phi; a_0, \phi_0) = \sum_{n} {\cal{K}}_{n} (a ; a_0) \; K_n ( \phi; \phi_0) 
\label{ker}
\ee
The matter kernel $K_n ( \phi; \phi_0)$ is equal to $\scal{\phi}{\psi_n}\scal{\psi_n}{\phi_0}$
as in Schr\"odinger settings. The gravitational kernel ${\cal{K}}_{n} (a ; a_0)$ is a solution
of eq. (\ref{M12}) and satisfies specific boundary conditions, see \cite{wdwgf}.
In the WKB approximation, for $a>a_0$, it is equal to
\be
{\cal{K}}_{n}(a; a_0) =  \Psi_{W\!K\!B} ( a; n) \Psi_{W\!K\!B}^*( a_0; n )
\label{kerkb}
\ee
At this point, we wish to point out that we used a dissymmetrical
treatment of $a$ and $\phi$ when constructing the kernel ${\cal{ K}}$
or expressing the general solution of the WDW equation in eq. (\ref{M13}).
As in Schr\"odinger settings, only the states of the matter field $\phi$ 
were quantized, i.e. imposed to belong to a 
Hilbert space.
This dissymmetry is reinforced upon considering many
matter fields since only $a$ will not participate to the definition of the Hilbert space.

When the field configurations are such that the dominant contribution
comes from states with energy distributed around $E_{\bar n}(a)$,
one can expand the gravitational kernels around that energy.
For a discussion concerning the validity of this expansion, see \cite{wdwgf}.
By using eqs. (\ref{M8}, \ref{M144}) and by developing the phase\footnote{
It should be noted that one deals with a first order expansion in the {\it phase}.
The kernel itself is still a non-linear function of  $E_n - E_{\bar n}$
since one keeps all terms of the exponential. This is how the classical behavior is recovered 
from quantum mechanics\cite{mtw}.}
to first order in $E_n - E_{\bar n}$, one obtains
\ba
{\cal{K}}(a, \phi; a_0, \phi_0) &=& \Psi^*( a_0;\bar  n )\Psi ( a;\bar  n )
\nonumber\\&& 
\times \sum_{n}  exp\left[{-i\int_{a_0}^a da' 
{a \over G p_{\bar n}(a')}
 [E_n(a') - E_{\bar n}(a')]}\right]
K_n ( \phi; \phi_0) 
\nonumber\\&=& 
\Psi^*( a_0;\bar  n )\Psi ( a;\bar  n )
 \sum_{n}  exp\left[{-i\int_{0}^{t_n} dt' [E_n(t') - E_{\bar n}(t')]}\right]
K_n ( \phi; \phi_0) \nonumber\\&=& 
\Psi^*( a_0;\bar  n )\Psi ( a;\bar  n )\ e^{-i\int_{0}^{t_n} dt'  E_{\bar n}(t')} \times K ( \phi; \phi_0, {t_n})
\label{kern2}
\ea
In the second line, we have introduced the Hamilton-Jacobi time, 
the conjugate to the matter energy $E$,
defined by
\be
\Delta  t_{\bar n}(a) = \partial_E \int^a_{a_0} da' p(a', E) \vert_{E=E_{\bar n}}
 = \int^a_{a_0} da' {a' \over G p_{\bar n}(a')}= \int^a_{a_0} da' { 1 \over \dot a_{\bar n} (a')}
\label{hjt}
\ee
By definition it equals the proper time evaluated in the geometry driven by $E_{\bar n}$
and given in eq. (\ref{dota}).
In the third line of eq. (\ref{kern2}), we have made use of the relation between
the time dependent kernel $K ( \phi; \phi_0, t)$ and its decomposition in terms
of eigenstates of given energy.

Eq. (\ref{kern2}) shows that the kernel ${\cal{K}}$ to propagate from $a_0, \phi_0$ to $a, \phi$ 
can be decomposed in a purely gravitational term controlled by the 
mean energy times the usual time dependent matter kernel $K$ evaluated in the ``mean''
geometry, i.e. in the geometry driven by the mean energy $E_{\bar n}$. 
Notice that this (kinematical) time dependence was obtained in two steps.
Firstly, through a first order change of the gravitational action with respect to the
matter energy $E_n$ and secondly, by using the ``dispersion relation'' $
\partial_{E} p (a, E) = 1/\dot a$ (i.e. the Hamilton equation $dE/dp = \dot a$) stemming from the
dynamical character of gravity. In view of both steps, there is a the strong
analogy between this development leading to a time dependent kernel 
and the development that leads to the concept
of a canonical partition function of a little system contained in a bigger ensemble
which is microcanonically distributed, see App. B in \cite{wdwgf}. 
In each case, it is through a first order change in the 
partitioning energy that the ``classical'' concept of time (or inverse temperature) is determined
from the properties of the heavy system.
Moreover, in both cases, the physical justification of considering
fluctuations of the partitioning energy arises from the existence
of interactions, see next subsections.

From eq. (\ref{kern2}), it is straightforward to make contact with classical mechanics.
To this end, one should 
suppose that the matter kernel can also be correctly approximated 
by its WKB expression. In this case, it is entirely
 governed by the matter action $S=\int^\phi_{\phi_0} d\phi' \pi(\phi', E_n)$
where $\pi(\phi, E_n)$ is the momentum of $\phi$ at fixed $E_n$.
Then, the dominant contribution to ${\cal{K}}(a, \phi; a_0, \phi_0)$ 
comes from energy repartitions located near the 
saddle point of its phase. The location of this saddle point is given by the
solution of 
\ba
\partial_{E_n}\left[ \int^a_{a_0}\!da' p(a', E_n) -  
\int^\phi_{\phi_0}  \!d\phi' \pi(\phi', E_n)\right]&=& 0
\nonumber\\
\Delta t_{\bar n}(a, a_0) - \int^\phi_{\phi_0} d\phi' {1\over\dot\phi_{\bar n}(\phi')} &=& 0
\label{phidot}
\ea
In the second equality, we have used the dispersion relation for the matter field to rewrite
$\partial_{E} \pi(\phi, E)\vert_{E=E_{\bar n}}$ 
by $1/\dot \phi_{\bar n}$ exactly as we just did with $a$. 
Eq. (\ref{phidot}) means that the saddle value
 $E_{\bar n}$
is such that the 
lapses of time evaluated separately for gravity and matter agree. 
This constructive interference condition (see Box 25.3 in \cite{mtw}) can be 
conceived as the ``dual'' of a resonance condition in the following
sense. In traditional time dependent settings, the dominant 
contribution to quantum processes arises from states such that the energy is 
conserved, {\it c.f.} the Golden Rule\cite{LL}.
Here, in quantum cosmology, energy conservation is built in, thanks to the constraint
equation. 
Thus, the phases of sub-systems interfere constructively 
such that their classical times agree. 
This is exactly like the zeroth law of thermodynamics: 
at equilibrium, (inverse) temperatures agree.

In physical terms, eq. (\ref{phidot}) means that the cosmological time 
$\Delta t_{\bar n}(a, a_0)$ extracted from the
expansion law (which has a similar status to that of the ``ephemeris'' time based on
the solar system dynamics) is equal to the ``cesium'' time obtained from the 
(microscopic) behavior of matter. (This equality is not a tautology
since it relates uncoupled dynamical systems characterized by widely
separated time scales.)
Notice finally that this condition of equal times
 can be formulated in purely classical terms. 
Indeed it provides the answer to the following question\cite{bprmp}:
 Given $a_0, \phi_0$, what is the value of $E_n$ such that 
 $\phi$ is reached at $a$ ?
Thus we have established that the emergence of time in the
quantum kernel ${\cal{K}}$, see eq. (\ref{kern2}), makes use of classical 
concepts only: As in eq. (\ref{phidot}),  it follows, through a
first order variation of $E$,
from the classical relationship $\partial_{E} p (a, E) = 1/\dot a$.

The second point that we wish to address concerns the relationship between 
the exact solutions of eq. (\ref{M14}) and their WKB approximate expressions, eq. (\ref{M144}).
This relationship  is needed to properly investigate the consequences
of quantizing the propagation of $a$ expressed by eq. (\ref{wdw}). Had we obtained 
a first order equation in $\partial_a$, this would have meant that $a$ had no dynamics
at all, like the longitudinal part of the electric field in classical electrodynamics (or in QED)
which is fully determined by the charge density (operator).
Being second order, eq. (\ref{wdw}) implies that some backscattering might, and in general will,
be spontaneously generated. 
This quantum effect cannot be expressed in terms of matter states unlike
the Coulomb-Coulomb interaction which can be represented by a composite
operator of charged fields. 
Moreover, gravitational 
backscattering will modify the propagation of matter states as we shall see below.

To express the exact solutions of eq. (\ref{M14}) in terms of their WKB expressions,
we follow the usual technique of replacing a second order equation by a set of
two coupled first order ones, see \cite{wdwpc} for more details.
The exact solution is first decomposed as
\be
\Psi(a;n) = {\cal{C}}_n(a) \Psi_{W\!K\!B} (a; n) + 
{\cal{D}}_n(a) \Psi^*_{W\!K\!B} (a; n)\label{wkbexp}\ee 
and the   
the coefficients ${\cal{C}}_n(a)$ and ${\cal{D}}_n(a)$
are fully determined by requiring that
$\partial_a \Psi(a;n)$ be instantaneously decomposable into purely forward and
backward waves 
\be
i \partial_a \Psi(a;n) = p_n(a) \left[  {\cal{C}}_n(a) \Psi_{W\!K\!B} (a; n) -
{\cal{D}}_n(a) \Psi^*_{W\!K\!B} (a; n)\right]
\label{OTHER}\ee
This guarantees that ${\cal{C}}_n(a)$ and ${\cal{D}}_n(a)$
 are constant in the 
adiabatic limit $\partial_a p_n/p_n^2 \to 0$. In addition,
the Wronskian takes the simple form
\be
 W_n = \Psi^*( a; n)\ 
i\!\!\lr{\partial_{a}} \Psi ( a;n ) = \vert{\cal{C}}_n(a) \vert^2 - \vert {\cal{D}}_n(a)  \vert^2 
=constant
\ee
Simple algebra then yields the coupled first order equations 
\ba
\partial_a {\cal{C}}_n(a) &=&
{1\over 2} {\partial_a p_n(a) \over p_n(a) }\; e^{-2i\int^a da' p_n(a')}\;{\cal{D}}_n(a)
\nonumber\\
\partial_a {\cal{D}}_n(a) &=&
{1\over 2} {\partial_a p_n(a) \over p_n(a) }\; e^{2i\int^a da' p_n(a')}\; {\cal{C}}_n(a)
\label{1WKB}
\ea
These equations are equivalent to the original equation for $\Psi(a;n)$,
eq. (\ref{M14}).
They constitute a convenient starting point for evaluating perturbatively 
 non-adiabatic transitions from ${\cal{C}}_n(a)$ to ${\cal{D}}_n(a)$
(i.e. backscattering).
Moreover they resemble to the Schr\"odinger equation (\ref{eqtcn2})
that governs non-adiabatic transition in the particular case
of two eigenstates with $\scal{\partial_a\psi_m(a)}{ \psi_n(a)}$
 replaced by $\partial_a p_n(a)/p_n(a)$, see \cite{wdwpc} for a 
more detailed comparison.

\vskip.5 truecm
%
{\bf{The double adiabatic treatment}}
\vskip.2 truecm
\noindent
In this subsection, we consider the non-degenerate cases
in which the eigenstates of the matter hamiltonian depend on
$a$.
In these cases, the coefficients
${\cal{C}}_n$ also depend on $a$, like the $c_n(a)$ in eq. (\ref{eqtcn2}).
To obtain the equation which governs their evolution, we need to join
the usual adiabatic treatment
presented in eqs. (\ref{insteig}-\ref{eqtcn2})
with the treatment by wich eq. (\ref{M14}) is represented by eqs. (\ref{1WKB}).

To this end, we first carry out the instantaneous 
diagonalization of $H_M$, exactly like in eq. (\ref{insteig}).
We emphasize that this diagonalization does not
require the ``existence'' of a Schr\" odinger equation. 
Using these instantaneous eigenstates,
$\ket{\Xi(a)}$, solution of 
eq. (\ref{wdw}),
can always be decomposed as 
\be
\ket{\Xi(a)}
= \sum_n \varphi_n(a)
\ket{\psi_n(a)}
\label{decomp1}
\ee
The novelty of this decomposition with respect to eq. (\ref{M13}) is 
that the waves $\varphi_n(a)$ are no longer 
decorrelated since the instantaneous eigenstates
$\ket{\psi_n(a)}$ do now depend on $a$.
The difference 
with the adiabatic treatment of section 2
 arises from the fact that the WDW constraint is
second order in $\partial_a$. Therefore, to obtain a first order equation
for the coefficients ${\cal{C}}_n$, we must proceed 
to a second adiabatic development.
Thus, as in eq. (\ref{wkbexp}), we express $\varphi_n(a)$ in terms of the 
WKB waves 
\be
\varphi_n(a) = {\cal{C}}_{n}(a)\ \Psi_{W\!K\!B} (a; n)
+  {\cal{D}}_{n}(a) \ \Psi_{W\!K\!B}^* (a; n)
\label{decomp22}
\ee
In view of the coupling among eigenstates, we must generalize 
eq. (\ref{OTHER}). We require now that
\be
\bra{\psi_n(a)} i\!\! \stackrel{\rightarrow} \partial_a \!\ket{ \Xi(a) }= p_n(a)
\left[ {\cal{C}}_{n}(a)\ \Psi_{W\!K\!B} (a; n) -  {\cal{D}}_{n}(a)\ \Psi^*_{W\!K\!B} (a; n) 
\right]
\label{AUXWKB}\ee  
This equation still expresses that $\partial_a \ket{ \Xi(a)}$ is instantaneously 
decomposed into a superposition of forward and backward waves.
With this condition,
 the total current carried by  $\ket{\Xi(a)}$
contains no terms proportional to $\partial_a {\cal{C}}_{n}(a)$.
Indeed, one finds
\be
\bk {\Xi(a) \vert i\!\!\lr{\partial_{a}} \vert \Xi (a)} = \int d\phi \left[ \Xi^*(a, \phi)
i\!\!\lr{\partial_{a}} \Xi (a, \phi) \right]
= \sum_n \vert {\cal{C}}_{n}(a) \vert^2 - \sum_n
\vert {\cal{D}}_{n}(a) \vert^2 = C
\label{CONSCURR}\ee
Notice that the absence of relative factors in the above sum
 follows from the usual choice of working with equally normalized 
eigenstates, eq. (\ref{normal}), as well as our choice of identical (unit) Wronskians
for the WKB waves that form our gravitational basis.

Finally, by generalizing the derivation of eq. (\ref{1WKB}), one finds\cite{wdwpc}
\ba
\partial_a {\cal{C}}_{n} &=& \sum_{m \neq n}
\scal{\partial_a \psi_m}{\psi_n}
{p_n + p_m\over 2\sqrt{p_n p_m}}
\; e^{-i \int ^a (p_n - p_m) da }
\; {\cal{C}}_{m}\nonumber\\
& & + \sum_m \scal{\partial_a \psi_m}{\psi_n}
{p_n - p_m\over 2\sqrt{p_n p_m}}
\; e^{-i \int ^a (p_n + p_m) da }\; 
{\cal{D}}_{m}\nonumber\\
& &+{\partial_a p_n \over 2 p_n} 
e^{-2i \int ^a p_n da } \;  {\cal{D}}_{n} 
\label{central}
\ea
and the same equation with ${\cal{C}}_{n} \leftrightarrow {\cal{D}}_{n}$,
$i \leftrightarrow -i$. These equations are equivalent to the original
WDW equation (\ref{wdw}), exactly like eq. (\ref{eqtcn2}) is just a rewriting 
of eq. (\ref{sch}).
Moreover they furnish a very convenient starting point for 
answering the three questions raised in the Introduction.
We now turn to these aspects as well as
those concerning the interpretation of $\ket{\Xi(a)}$ in the light of 
the properties of the ${\cal{C}}_n(a)$.

\vskip.5 truecm
{\bf{The (physical) notion of evolution from non-adiabatic transitions}}
\vskip.2 truecm
\noindent
In the limit where both the adiabatic approximation for the matter states
and the WKB approximation for gravity are valid, 
the coefficients ${\cal{C}}_{n}$ and ${\cal{D}}_{n}$ are 
constants. In this case, we recover the (preparatory) situation of eq. (\ref{M13})  
wherein there is neither correlations among the coefficients 
 ${\cal{C}}_{n}$ and nor correlations with the ${\cal{D}}_{n}$.
One simply obtains a frozen superposition of {\it uncorrelated}
eigenstates
with $a$-dependent kinematical phase factors.

To given physical meaning to these phases, one must consider
non-adiabatic transitions. 
The simplest case consists in working in the adiabatic approximation for gravity only.
Then, one can neglect the coupling between
the coefficients ${\cal{C}}_{n}$ and ${\cal{D}}_{n}$. The simplified dynamical equation is
\be
\partial_a {\cal{C}}_{n} = \sum_{m \neq n}
\scal{\partial_a \psi_m}{\psi_n}
{p_n + p_m\over 2\sqrt{p_n p_m}}\;
e^{-i \int ^a (p_n - p_m) da' }\;
{\cal{C}}_{m}(a)
\label{schr'}
\ee
which implies immediately 
\be
 \sum_n \vert {\cal{C}}_{n}(a) \vert^2 = constant
\label{unit}
\ee

If one further assumes that the ${\cal{C}}_{n}$ form a well defined wave packet
 in $n$ centered around the mean $\bar n$, 
one can develop eq. (\ref{schr'}) around that mean, in power
of $n - \bar n$. 
To first order in $n - \bar n$, one obtains 
\ba
\partial_{a} {\cal{C}}_{n} &=& \sum_{m \neq n} \scal{\partial_{a} \psi_m}{\psi_n}
\; exp \left[ -i \int^a da' { 1 \over \dot a_{\bar n}(a') } \left\{ E_n(a') - E_m(a')\right\} \right]
{\cal{C}}_{m}(a)
\nonumber\\
&=&\sum_{m \neq n} \scal{\partial_{a} \psi_m}{\psi_n}
\; exp \left[-i \int^{t_{\bar n}(a)} dt'  \left\{E_n(a_{\bar n}(t')) - E_m(a_{\bar n}(t'))
\right\}\right]
{\cal{C}}_{m}(a)\quad\quad
\label{centralS}
\ea
This equation is {\it identical} to eqs. (\ref{eqtcn}, \ref{eqtcn2}). 
More precisely, through $dt_{\bar n}/da = 1/\dot a_{\bar n}(a)$,
i.e. the (inverse) expansion {\it rate} of the mean universe,
we recover the Schr\"odinger equation governing
non adiabatic transitions among instantaneous matter states
if one identifies the coefficients ${\cal{C}}_{n}(a)$
with the probability amplitudes $c_n(a)$ to find the $n-th$ state at $a$ in 
this mean universe.
We emphasize the {\it a posteriori} character of this identification.
Indeed, it is based on the comparison of eq. (\ref{eqtcn2})
with the resulting simplified equation governing the ${\cal{C}}_{n}(a)$.
Notice also that the justification of developing the expression
around the mean value $\bar n$ follows from a closer examination
of higher order terms in $n- \bar n$, see \cite{wdwgf, wdwpt}.
The mathematical criterion that legitimizes a first order expansion
in $n-\bar n$ is that the spread in energy satisfies $\bk{(E_n - E_{\bar n})^2}
<\!\!< E_{\bar n}^2$. This requires that the universe be 
macroscopic\cite{vil}\cite{BV}\cite{wdwgf}. Notice that
the Planck mass does not appear in the mathematical criterion.

When the ${\cal{C}}_{n}$ do not form a well defined wave packet, it is
meaningless to expand in power of $n - \bar n$ around some value
and to use the Hamilton-Jacobi time characterizing the corresponding geometry.
Instead, one should work directly with eq. (\ref{schr'}) and keep $a$ as the parameter
to follow the evolution of the ${\cal{C}}_{n}(a)$. Then,
even though each state ``lives'' in its geometry, i.e. is entangled to its
gravitational wave, it is still 
mandatory to interpret the weight ${\cal{C}}_{n}(a)$ as the probability amplitude
to find the $n$-matter state at $a$. Indeed, the three properties used by 
Max Born and mentioned after eq. (\ref{dota}) are still found. 

Once this identification is done, the interpretation of the full wave
function $\ket{\Xi (a)}$ is determined. In particular, due to the
second order character of the WDW constraint, the norm of  $\ket{\Xi (a)}$
does not determine probabilities since 
\be
\bk{\Xi(a) \vert \Xi(a)} = \sum_n \bk{\Xi(a) \vert \psi_n(a)}\bk{\psi_n(a)\vert \Xi(a)}
 =\sum_n {\vert {\cal{C}}_{n}(a)\vert^2 \over  p_n(a)} \neq constant
\label{notp}
\ee 
Instead the current leads to the correct expression, see eqs. (\ref{CONSCURR}, \ref{unit}).
That the norm of $\ket{\Xi (a)}$ does not determine probabilities 
is free of physical 
consequences, at least in the present case where there is no 
coupling among ${\cal{C}}_{n}(a)$ and ${\cal{D}}_{n}(a)$.

Before considering the problems that arise when these couplings
are taken into account, we wish to 
emphasize that the properties we just discussed, namely the 
recovery of the Schr\"odinger equation and the fact that 
the norm of $\ket{\Xi (a)}$ does not determine probabilities, are
not specific to cosmology. Indeed they are also found upon
considering the BFA limit in other dynamical models,
such as accelerated particle detectors or mirrors\cite{rec, Bfa}.

\vskip.5 truecm
{\bf{The 
difficulties induced by the ${\cal{C}}_{n}-{\cal{D}}_{n}$
 couplings}}
\vskip.2 truecm

\noindent
The last two terms in eq. (\ref{central}) govern
the couplings between the ${\cal{C}}_{n}(a)$ and the ${\cal{D}}_{m}(a)$.
Once they are taken into account, they induce quantum transitions
from ${\cal{C}}_{n}\ket{\psi_n(a)}$ to ${\cal{D}}_{m}\ket{\psi_m(a)}$. 
Mathematically, these transitions imply that the 
knowledge of the initial values of all ${\cal{D}}_{m}$
is required in order to determine the evolution of the ${\cal{C}}_{n}$.
Physically, it means that matter states associated with 
expanding and contracting universes are quantum mechanically 
interacting. In other words, the matter states in
our expanding universe do not form (exactly) 
a closed set, as if our universe were not isolated.

To our opinion, there is no satisfactory proposition which
specifies how to deal with this problem. The main difficulties are:

\noindent
i. how to cope with the conservation of the Wronskian, eq. (\ref{CONSCURR}),
that is the 
violation of eq. (\ref{unit}) ?

\noindent
ii. how to determine the values of the ${\cal{D}}_{m}$ which are 
relevant for early cosmology ?

We emphasize the double aspect of the problem. There is both a 
conceptual side, i.e. to find the appropriate framework of interpretation, and a 
more pragmatic aspect which concerns the actual values of the ${\cal{D}}_{m}$
and their relevance for our universe.

Various propositions have been made in order to try to cope with
this double problem. All propositions somehow fall
 within one of the following four classes.
We refer to \cite{isham} for a detailed comparative description.
In what follows, we shall briefly present what constitutes, 
to our opinion, their main weakness.

1. A first approach consists in rejecting the WDW equation itself since its
solutions do not have a simple satisfactory interpretation. Then the main difficulty 
is to find a principle that selects the new equation. In particular, this principle
should explain why the new equation is not the quantized version of 
the Hamilton-Jacobi constraint equation 
which is quadratic in $\partial_a S_{gravity}$. 
We recall that the canonical quantization of such a quadratic term
inevitably generates backscattering effects that caused the
abandonment of the WDW equation. 

2. A second one consists in searching for a new definition of probabilities
no longer based on the current carried by $\ket{\Xi (a)}$.
The so-called ``conditional probabilities'' constitute an example.
In this case, the main difficulty is to explain why it is physically meaningful
to consider probabilities which contain superpositions of 
expanding and contracting universes of the type given in eq. (\ref{decomp22}). 
What are the physical questions whose answers are sensitive 
either to an interfering term ${\cal{C}}_n {\cal{D}}_m^*$ or even
to a {\it sum} like $\vert {\cal{C}}_n \vert^2 +\vert {\cal{D}}_n \vert^2$ ?
Moreover, we recall that the ``conditional probability'' to find matter
in the $n$-state depends on $a$ even in the 
absence of transition\cite{wdwin}.

3. Third quantization offers an alternative method to deal with ${\Xi (a, \phi)}$.
In this perfectly consistent framework, backscattering is interpreted as 
pair creation of (macroscopic) universes. 
However, additional principles are required to define the ``vacuum'',
 to pick the initial state as well as to be able to answer questions 
of the type: Given the initial state, 
what is the probability to find, at $a$,
 a specific matter state in the expanding sector ?
Without these additional principles, third quantization is rather useless.

4. The fourth attitude consists in admitting that probabilities and/or
unitarity are intrinsically approximated concepts when restricted to 
the sole expanding sector of the theory.
By adopting this attitude, one simply
 postpones confronting the problem of the
${\cal{C}}_{n}-{\cal{D}}_{n}$ couplings.


\vskip.5 truecm
{\bf{Conclusions}}
\vskip.2 truecm
\noindent

In this article, we have applied a double adiabatic treatment to the
solutions of the WDW equation, eq. (\ref{wdw}).
The power of this treatment is displayed by its re-expression
given in eq. (\ref{central}).
In itself, this rewriting does not introduce any kind of approximations.
What it does instead is to deliver the precise nature of the approximations
that reduces the WDW equation to the corresponding Schr\"odinger equation.
These approximations are the following: One should neglect the coupling between
expanding and contracting solutions in order to obtain a monotonic expansion 
and secondly, one must proceed to a first order expansion in the matter energy 
spread. No less no more.

Moreover, upon neglecting only the coupling between
expanding and contracting solutions, one still obtains
a unitary evolution even though it is not controlled by a time parameter,
i.e. it cannot be obtained from a Schr\"odinger equation. Indeed, the 
evolution should be followed by $a$ itself.

Both simple results directly follow from the 
fact that the adiabatic treatment has been applied to the WDW equation.
Indeed it
re-organizes the dynamical variables so as to 
express the content of quantum cosmology (rather than 
quantum gravity in full generality), i.e.
the dynamical interplay between the expansion of the universe
described by the (homogeneous part of) the conformal factor $a$ and 
(local\cite{wdwgf}) matter fields.
Had we used time dependent perturbation theory, no
simple characterization of the approximations that should be implemented
to recover the Schr\"odinger equation would have been obtained.
The reason is that there is an interplay between higher order terms in
the coupling constant and the gravitational distortion introduced by the
level shifts, see \cite{wdwpt}. The adiabatic treatment sorts out these
effects by construction.

We wish to emphasize this last point: Due to the dynamical coupling 
between matter and gravity, the concept of 
{\it the} Schr\"odinger equation (expressable in any representation,
{\it c.f.} eq. (\ref{sch}) and eq. (\ref{eqtcn}))
no longer exists in quantum cosmology.
This is particularly clear upon considering matter fields carrying
non-vanishing momenta since the coupling to gravity occurs through
energy. In that case indeed, the representation
of Fock states in terms of local configurations, i.e. {\it not} characterized by a 
given energy-momentum, is a secondary approximate 
concept, see App. A of \cite{wdwgf}.

\vskip .3 truecm

{\bf Acknowlegdments}

\noindent
I am very much indebted to Serge Massar for the 
clarifying discussions concerning 
the physical relevance of 
adiabatic treatments.

\end{document}